\documentclass[aps,twocolumn,pra,showpacs]{revtex4}

\usepackage{graphicx,amsmath,amssymb}
\usepackage{color}

\begin{document}

\title{Spatial and temporal characterization of a Bessel beam produced using a conical mirror}

\author{K. B. Kuntz,$^1$ B. Braverman,$^2$ S. H. Youn,$^{1,3}$ M. Lobino,$^1$ E. M. Pessina,$^4$ A. I. Lvovsky$^{1,}$}
\email{lvov@ucalgary.ca}

\affiliation{$^1$ Institute for Quantum Information Science,
University of Calgary, Calgary, Alberta T2N 1N4, Canada}

\affiliation{$^2$ Department of Physics, University of Toronto,
Toronto, Ontario M5S 1A7, Canada}

\affiliation{$^3$ Department of  Physics, Chonnam National
University, Gwangju, 500-757, Korea}

\affiliation{$^4$ Pirelli Labs, Viale Sarca 336, 20126 Milano,
Italy}

\begin{abstract}
We experimentally analyze a Bessel beam produced with a conical
mirror, paying particular attention to its superluminal and
diffraction-free properties. We spatially characterized the beam in
the radial and on-axis dimensions, and verified that the central
peak does not spread over a propagation distance of $73$
$\mathrm{cm}$. In addition, we measured the superluminal phase and
group velocities of the beam in free space. Both spatial and
temporal measurements show good agreement with the theoretical
predictions.
\end{abstract}

\pacs{42.25.Bs}
\maketitle

\section{Introduction}

A special class of non-diffracting light waves called Bessel beams
\cite{Durnin4} have been extensivelly studied over the last two
decades. Bessel beams are characterized by a transverse field
profile in the form of a zero-order Bessel function of the first
kind. They exhibit several intriguing properties, such as
diffraction-free propagation of the central peak over a distance
fixed only by the geometry of the source device, and superluminal
phase and group velocities in free space. Due to their
diffraction-free characteristics, they have found applications in
several fields of physics. Non-diffractive pump fields have been
utilized in nonlinear optical processes like parametric down
conversion \cite{BelviKazak,Belyi,Orloy,Trapani,Longhi,Piskarskas},
second \cite{Wulle,Arlt,Ding,Jedrkiewicz}, third \cite{Peet}, and
higher-order harmonic generation \cite{Averchi}. Furthermore, the
accuracy of optical tweezers \cite{Milne,Garces,Summers}, optical
trapping \cite{Arlt63,Garces66,Fan,Tatarkova,Arakelyan} and the
resolution of medical imaging \cite{Rolland,Jianyu} have been
enhanced by the implementation of diffraction-free beams.

In addition to exploiting the non-diffractive feature of Bessel
beams, superluminality has also been a subject of investigation.
There have been attempts to demonstrate this property using light in
the microwave range \cite{Mugnai} but the experimental uncertainties
induced by the apparatus were large compared to the magnitude of the
superluminal effect \cite{Ringermacher,Bigelow}. Recently, the
propagation velocity of an ionization wavefront induced by a Bessel
pulse has also been measured \cite{Alexeev}, but until now no
measurement revealed both superluminal group and phase velocities in
free space.

Here, we report a complete characterization of spatio-temporal
properties of an optical Bessel beam generated by reflection from a
conical mirror \cite{Fortin}. This novel technique of producing
Bessel beams has the advantage, with respect to commonly used
axicons \cite{Mcleod}, of avoiding dispersion and is thus more
suited for applications that require ultrashort laser pulses.

In the following we describe the spatially characterization of the
beam, showing non-diffractive properties over its propagation range.
Later we show a set of measurements of superluminal phase and group
velocities in free space that has greater accuracy than ever
previously measured.

\section{Beam Properties}

A non-diffractive Bessel beam is realized by a coherent
superposition of equal-amplitude, equal-phase plane waves whose
wavevectors form a constant angle $\theta$, called the axicon angle,
with the direction of propagation of the beam which we define as the
$z$-axis.

As originally proposed by Durnin \cite{Durnin4}, the analytical
solution of the Helmholtz equation for the electric field propagation is given by:
\begin{equation} \label{solution}
    E(r,z,t) = A \exp\left[i(\beta z-\omega t)\right] J_0(\alpha r)
\end{equation}
where $A$ is constant, $\beta=k\cos\theta$,
$\alpha=k\sin\theta$, $k=\omega/c$ is the wave-vector,
$\omega$ is the angular frequency, $r$ is the radial coordinate and
$J_0$ is a zeroth-order Bessel function of the first kind. Equation
(\ref{solution}) defines a non-diffracting beam since the
intensity distribution is independent of $z$ and equal to
$J_0^2(\alpha r)$. Furthermore, the field described in Eq.
(\ref{solution}) propagates with a superluminal phase velocity given
by $v_p=c/\cos\theta>c$. The independence of the phase velocity
from the field frequency implies that in free space the group
velocity is also superluminal and is equal to the phase velocity.

\begin{figure}[b]
  \center{\includegraphics{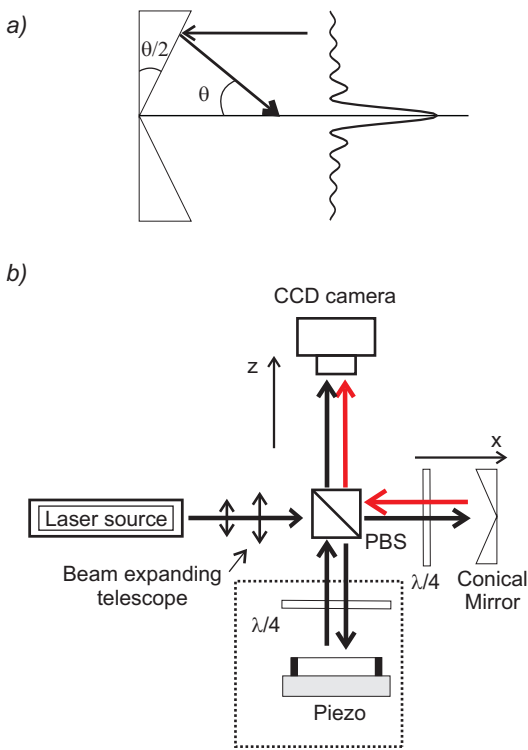}}
  \caption{(color online). a) Conical mirror used to produce the Bessel beam.
  b) Schematic of the experimental setup used for the spatial and temporal characterization of a Bessel
  beam. The elements added for temporal characterization are inside a dotted square. Black arrows denote the Gaussian beam path while red arrows
  refer to the Bessel beam path. PBS, polarizing beamsplitter, $\lambda/4$, quarter waveplate.}
  \label{fig:1}
\end{figure}

In practical experiments the diffraction free region is limited by
the finite extent of the beam. As evidenced by Fig. \ref{fig:1}(a),
the maximum diffraction free propagation distance depends on the
radius $R$ of the optical element used for producing the beam and on
the axicon angle $\theta$, and is given by $z_{max}=R/\tan\theta$.
At propagation distances greater than $z_{max}$, the beam diffracts
very quickly, spreading the energy over an annular region.

In our experiment, the optical element used to generate the Bessel
beam is a conical mirror with a radius of $1.27$ $\mathrm{cm}$ and
an apex angle of $\pi-\theta=179^\circ$. The mirror parameters
correspond to a non-diffractive distance of $73$ $\mathrm{cm}$ and
lead to the phase and group velocities exceeding $c$ by $0.015\%$.

\section{Spatial Characterization}

The non-diffractive propagation was verified by spatial
characterization of the Bessel beam. The experimental setup used for
these measurements is shown in Fig. \ref{fig:1}(b). The light source
is a continuous wave (CW) He:Ne laser operating at $543.5$
$\mathrm{nm}$. The horizontally polarized beam is first expanded so
that the resulting collimated light has a waist slightly larger than
the diameter of the conical mirror, and then is transmitted through
a polarizing beamsplitter (PBS). After the PBS, the Gaussian beam
passes through a $\lambda/4$ waveplate, reflects off the conical
mirror, passes through the waveplate a second time, changing to a
vertical polarization, and is reflected by the PBS. In this way,
65\% of the Gaussian beam power is converted into a nondiffractive
beam.

\begin{figure}
  \center{\includegraphics{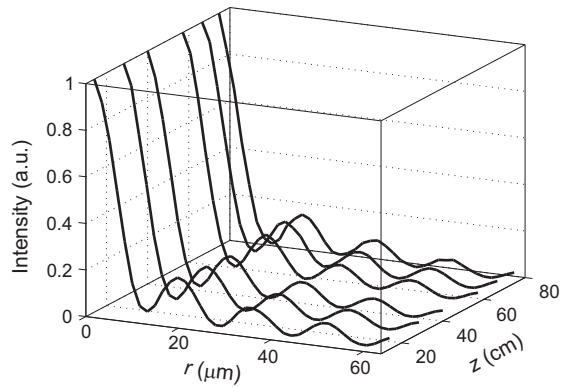}}
  \caption{(color online). Normalized radial intensity profile for several CCD
  camera positions.}
  \label{fig:3D}
\end{figure}

The intensity profile at various positions along the propagation
axis was recorded using a CCD camera. A microscope was constructed
to magnify the intensity profile by a factor of 3.5 so the camera
could resolve fine transverse features of the diffraction pattern.
The different radial intensity profiles recorded are shown in Fig.
\ref{fig:3D} and are almost identical over the entire diffraction
free region with a central peak width which is consistent with the
theoretical expectation of $(2\times 2.405)/\alpha=24 \pm1.5$
$\mathrm{\mu m}$, where 2.405 is the first zero of the Bessel
function.

\begin{figure}[t]
  \center{\includegraphics{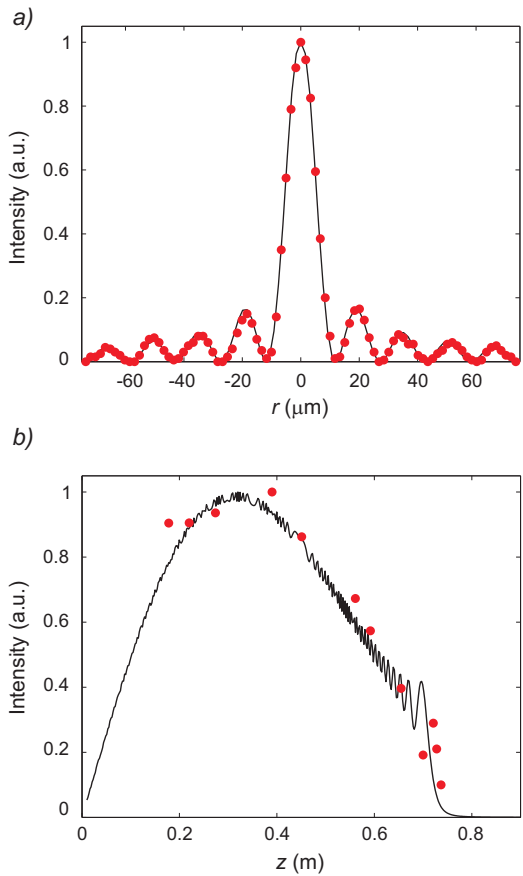}}
  \caption{(color online). a) Measured intensity profile (dots) and numerical calculation (line).
  b) On-axis Intensity: experimental data (dots) and theoretical calculation (line).
  }
  \label{fig:3}
\end{figure}

We compared the measured intensity pattern with the pattern
calculated numerically by the Rayleigh-Sommerfield diffraction
integral \cite{Goodman} (Fig. \ref{fig:3}(a)). For this calculation
we assumed the amplitude transmission function of the mirror is
given by:
\begin{equation} \label{transmission}
    T(r)=e^{-i2kr\tan(\theta/2)}.
\end{equation}
The intensity of the central peak as a function of the propagation
distance is shown in Fig. \ref{fig:3}(b) together with a the
numerical prediction. The central peak intensity behavior is
determined by the fact that the light that constructively interferes
to create the peak at a longitudinal position $z$ arrives from a
region of the mirror that has a circumference of $2\pi R=2\pi
z\tan\theta$. The intensity thus grows linearly for small values of
$z$ and then decreases when the intensity decrease of the Gaussian
profile of the input beam is more rapid then the linear increase of
the circumference. The oscillations at the end of the curve are due
to the Fresnel diffraction from the outer edge of the mirror.

The 73-cm non-diffractive distance estimated with geometrical optics
agreed with that obtained from the numerical calculation and with
the experimental results.

\section{Temporal Characterization}

The phase and group velocities in free space were measured in order
to verify that the propagation of the Bessel beam exceeds the speed
of light. We implemented an interferometric design using CW light
for the phase velocity measurement, and femtosecond pulses for the
group velocity measurement. The design, shown in Fig. \ref{fig:1}b,
is a mixed Michelson interferometer, with a plane mirror in one arm
and a conical mirror in the other. In the output channel of the
interferometer we placed a CCD camera to record the on-axis
intensity as a function of the camera's longitudinal position $z$.
We placed a $\lambda/4$ waveplate into each arm to permit
independent manipulation of the intensity of the two beams.

This arrangement allowed us to compare the group and phase
velocities of the Bessel beam relative to those of the Gaussian beam
(that are equal to $c$) and to observe the relatively small
superluminal effect expected.



\subsection{Phase Velocity Measurement}

Gaussian and Bessel beams have both the same frequency $\nu$ but
different phase velocities, $v_{p,G}$ and $v_{p,B}=v_{p,G}+\Delta
v_{p}$, respectively, along the $z$ direction. Thus they will show
interference fringes spaced at $\Delta z =v_{p,G}v_{p,B}/ (|\Delta
v_{p}|\nu)$. In our case, with $|\Delta v_{p}|<<v_{p,G}$,
$v_{p,B}\approx c$, we find $|\Delta v_{p}| = \lambda c/\Delta z$.

\begin{figure}[t]
  \center{\includegraphics{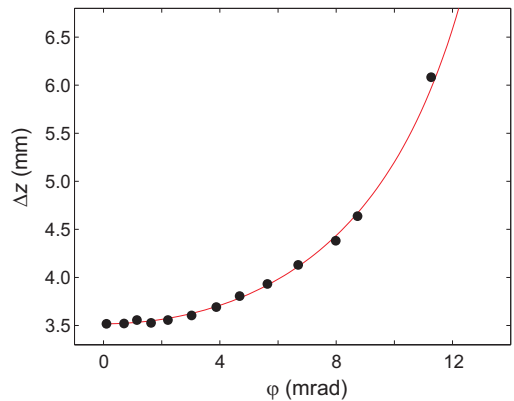}}
  \caption{(color online). CW measurements: period of the interference fringes $\Delta z$ as a function
  of the angle $\varphi$ between the direction of propagation of the Gaussian
  and Bessel beams (dots) together with the theoretical fit (line).}
  \label{fig:7}
\end{figure}

Measuring the spacing of the interference fringes thus yields the
magnitude of $\Delta v_{p}$. In order to determine the sign of
$\Delta v_{p}$, and hence whether the Bessel beam is propagating
superluminally, we varied the angle $\varphi$ between the direction
of propagation of the Gaussian and Bessel beam. By doing so, we set
the phase velocity component of the Gaussian beam in the direction
of propagation of the Bessel beam to
$v_{p,G}=\omega/k_z=\omega/k\cos\varphi=c/\cos\varphi$.
%
%
%
%
If the phase velocity of the Bessel beam exceeds that of the
Gaussian beam then with increasing $\varphi$ between
$0\leq\varphi<\theta$, the interference fringe spacing $\Delta z$
should also increase. Experimentally, we aligned the Gaussian beam
along the optical rail and manipulated $\varphi$ by slightly tilting
the conical mirror. We determined the value of $\varphi$ by
displacing the CCD along the optical rail and recording the relative
transverse positions of the centres of the two beams at each
location.
%
%
Figure \ref{fig:7} displays the measured interference period as a
function of $\varphi$ together with a theoretical fit in which the
axicon angle $\theta$ was the variable parameter. In particular,
note that with increasing misalignment, $\Delta z$ is increasing,
indicating that the phase velocity of the Bessel beam is higher than
that of the Gaussian beam.  The data give a phase velocity of
$v_{p}=(1.000155\pm0.000003)c$ which is consistent with the expected
value $c/\cos(\theta)=1.000152c$, corresponding to an axicon angle
of $1.009^\circ\pm0.01^\circ$.

\subsection{Group Velocity Measurement}

The group velocity measurement was performed using 200 fs pulses at
$\lambda = 776.9$ $\mathrm{nm}$ from a Ti:Sapphire mode-locked laser
pumped by a $532$ $\mathrm{nm}$ solid-state laser. The optics were
replaced to accommodate the new wavelength. The Gaussian and Bessel
beams were aligned with each other to within $0.5$ $\mathrm{mrad}$.

The intensity in the output channel of the interferometer was
measured with the CCD camera. When there is no temporal overlap
between the Gaussian and the Bessel pulse, the CCD measured the sum
of the intensities of the two pulses. On the other hand, when the
pulse arrival was simultaneous, the intensity was affected by
interference.
%
%
%
%
%
Suppose that for some configuration of the interferometer the CCD is
located in a position of maximum visibility. If the camera is now
translated along the propagation axis away from the PBS by a
distance of $\Delta z$, the interferometer arm with the planar
mirror must be moved by a distance $\Delta x$ in order to restore
the same visibility. Therefore, the Bessel pulse travels an extra
distance of $\Delta z$ in the time the Gaussian pulse travels a
distance of $\Delta z + 2\Delta x$, giving

\begin{equation}
    \Delta z=\frac{2v_{g,B}}{c-v_{g,B}}\Delta x.
    \label{eq:groupeq}
\end{equation}

\begin{figure}
  \center{\includegraphics{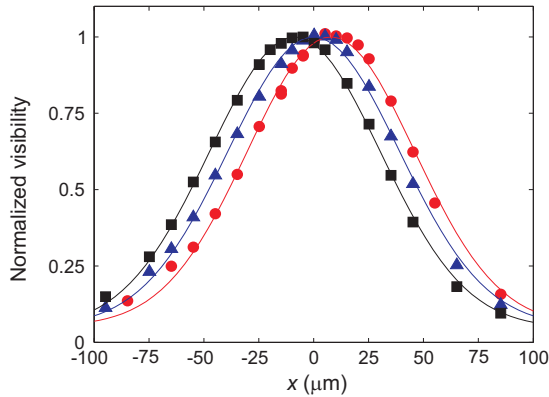}}
  \caption{(color online). Pulsed measurements: normalized fringe visibility for $z$ = 219($\scriptstyle{\blacksquare}$),
  119(\textcolor{blue}{$\blacktriangle$}) and 9(\textcolor{red}{$\bullet$}) mm, as a function of the
   plane mirror translation $\Delta x$. The experimental data are fitted with a Gaussian curve.}
  \label{fig:8}
\end{figure}

In the experiment, we acquired several pairs $(\Delta x, \Delta z)$
for which the interference visibility is optimized. The interference
was observed by scanning the plane mirror over a distance of $1-1.5$
wavelengths with a piezoelectric transducer. For a given position
$z$ of the camera, the interference visibility over a range of plane
mirror positions was evaluated (Fig. \ref{fig:8})

\begin{figure}
  \center{\includegraphics{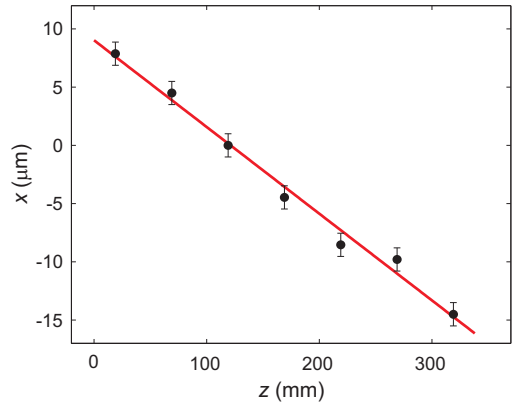}}
  \caption{(color online). Displacement $x$ of the Michelson interferometer plane mirror, at which the visibility is
  maximized, as a function of the CCD camera position $z$ (dots) with a linear fit (line).}
  \label{fig:9}
\end{figure}

The linear relationship predicted by  Eq. (\ref{eq:groupeq}) is
evidenced by Fig. \ref{fig:9}, and is characterized by a linear
regression slope $\partial x/\partial z =
(-7.5\pm0.3)\times10^{-5}$. The negativity of the slope indicates
that the relative path length of the Gaussian beam had to be reduced
as the overall travel distance increased (i.e. smaller $z$ values),
which corresponds to a superluminal group velocity, $v_{g} =
[1-2(\partial x/\partial z)]c = (1.000150\pm0.000006)c$. This
velocity corresponds to the apex angle of
$0.992^\circ\pm0.02^\circ$.

\section{Conclusion}

We have spatially and temporally characterized an optical Bessel beam
produced using a conical mirror propagating in
free space.

We ascertained the beam to have a constant peak size over the
propagation distance determined by the properties of the conical
mirror.

The phase and group velocities of the beam were determined with an
interferometric setup to be superluminal, with values of
$v_{p}=(1.000155\pm0.000003)c$ and $v_{g}=(1.000150\pm0.000006)c$,
respectively, which is consistent with the theoretical prediction.

\section*{Acknowledgements}
This work was supported by NSERC, CFI, AIF, Quantum$Works$ and CIFAR
(A.L.);

\end{document}